\documentclass[useAMS,usenatbib]{mn2e}
\usepackage{graphicx}
\usepackage{amssymb}
\usepackage[T1]{fontenc}

\title[Multifrequency detection of point sources in CMB maps]{A
  multifrequency method based on the Matched Multifilter for the
  detections of point sources in CMB maps} \author[Lanz et
  al.]{L.~F. Lanz$^{1,2}$\thanks{E-mail: lanz@ifca.unican.es},
  D. Herranz$^{1}$, J.~L. Sanz$^{1}$, J. Gonz\'alez-Nuevo$^{3}$ and
  M. L\'opez-Caniego$^{1,4}$ \\ $^{1}$ Instituto de F\'\i sica de
  Cantabria, CSIC-UC, Av. de Los Castros s/n, Santander, 39005, Spain
  \\ $^{2}$ Departamento de F\'\i sica Moderna, Universidad de
  Cantabria, Av. de Los Castros s/n, Santander, 39005, Spain \\ $^{3}$
  SISSA, via Beirut 4, I-34014 Trieste, Italy \\ $^{4}$ Astrophysics
  Group, Cavendish Laboratory, J.J. Thomson Avenue, CB9 0E1,
  Cambridge, United Kingdom}

\begin{document}

\date{Received --, Accepted --}

\pagerange{\pageref{firstpage}--\pageref{lastpage}} \pubyear{2009}

\maketitle

\label{firstpage}

\begin{abstract}

In this work we deal with the problem of simultaneous multifrequency
detection of extragalactic point sources in maps of the Cosmic
Microwave Background. We apply a linear filtering technique that uses
spatial information and the cross-power spectrum. To make this, we
simulate realistic and non-realistic flat patches of the sky at two
frequencies of \emph{Planck}: 44 and 100 GHz. We filter to detect and
estimate the point sources and compare this technique with the
monofrequency matched filter in terms of completeness, reliability,
flux and spectral index accuracy. The multifrequency method
outperforms the matched filter at the two frequencies and in all the
studied cases in the work.

\end{abstract}


\section{Introduction} \label{sec:intro}

Over the last few years, a big effort has been devoted to the problem
of detecting point sources in Cosmic Microwave Background (CMB)
experiments. The main reason is that modern CMB experiments have
reached resolution and sensitivity levels such that their capability
to estimate the statistics of CMB fluctuations at high multipoles is
no longer limited by instrumental noise but by Galactic and
extragalactic foreground contamination. Among extragalactic
contaminants, point sources are the most relevant, both in
temperature~\citep{tof98,zotti99,hob99,zotti05} and in
polarisation~\citep{tucci04,tucci05,powps}. Moreover, they are one of
the most difficult contaminants to deal with and, at least in the
frequency range spanned by CMB experiments, one of the most poorly
known. It is therefore mandatory to detect the maximum possible number
of extragalactic point sources (EPS) and to estimate their flux with
the lowest possible error before any serious attempt to study the CMB
anisotropies.

But EPS are not only a contaminant to get rid of. The study of EPS at
microwave frequencies is very interesting from the standpoint of
extragalactic astronomy. The next generation of CMB experiments will
allow one to obtain of all-sky EPS catalogues that will fill in the
existing observational gap in our knowledge of the Universe in the
frequency range from 20 to roughly 1000 GHz. We expect to derive
source number counts and spectral indices, to study source variability
and to discover rare objects such as inverted spectrum radio sources,
extreme gigahertz peaked spectrum sources (GPS) and high-redshift
dusty proto-spheroids~\citep[see for example the \emph{Planck
Bluebook,}][] {bluebook}. The first fruits of these new era of CMB
experiments are already here: the \emph{Wilkinson Microwave Anisotropy
Probe} (WMAP) satellite~\citep{wmap0} has made possible the obtaining
of the first all-sky point source catalogues above $\sim0.8$--$1$ Jy
in the 23--94 GHz range of frequencies in temperature
(\citet{wmap0,hinshaw07short,chen08,wright09short,NEWPS07,massardi09}
and polarisation~\citep{powps}, being complete the last three ones. In
a recent paper~\citet{gnuevo08} have shown how these catalogues can be
used to study the statistics of point sources in that range of
frequencies. The \emph{Planck} mission~\citep{planck_tauber05} will
allow us to extend these catalogues down to lower flux limits and up
to 857 GHz.

We have mentioned that the detection and estimation of the flux of EPS
are a difficult task. The main reason for this is that the many
different types of EPS that are distributed in the sky form a very
heterogeneous set of objects that do not have a common spectral
behaviour. While other foreground contaminants follow a specific
emission law that is approximately well known (or can be inferred from
observations) and that varies relatively slow and continuously across
the sky, each extragalactic point source has an emission law, that, in
principle, can be totally different to any other and independent from
them. From the point of view of statistical signal processing, the
problem of detecting EPS is a case of sorely under-determined
component separation problem where the number $M$ of components is
much larger than the number $N$ of frequency channels.

Let us consider the case of another contaminant that is of relevance
in CMB images and to which the EPS problem has some similarities: the
Sunyaev-Zel'dovich(SZ) effect~\citep{SZ70,SZ72,carlstrom02}. As in the
case of EPS, the SZ contamination occurs in small and compact regions
of the sky and it affects the high-$\ell$ region of the fluctuations
angular power spectrum. But in the case of the thermal SZ effect the
frequency dependence is very well known (ignoring relativistic
corrections). This has made possible the development of a number of
powerful detection techniques that are specifically suited to the
problem of SZ effect detection in CMB images
\citep{chema02,herr02a,hob03,pierpaoli05,herr05,vale06,pires06,melin06,bartlettSZ08,challenge08short}
or to use more generic diffuse component separation techniques
\citep[see for example, among many
others,][]{fastica02,vlad02,smica03,eriksen04,bedini05,bonaldi06} to
obtain SZ maps and catalogues.

Unfortunately, these approaches are not directly applicable to EPS
detection.  The most commonly used approach to this problem consists
on working separately in each channel. The key idea is to take
advantage of the fact that all the EPS have the same shape (basically,
that of the beam). In the field of CMB images, wavelet
techniques~\citep{vielva01,vielva03,MHW206,wsphere,NEWPS07}, matched
filters~\citep[MF,][]{tegmark98,barr03,can06} and other related linear
filtering
techniques~\citep{sanz01,naselsky02,herr02c,can04a,can05a,can05b} have
proved to be useful. All these techniques rely on the prior knowledge
that the sources have a distinctive spatial behaviour and this fact is
used to design some bandpass filter to enhance them with respect to
the noise. Detection can be further improved by including prior
information about the sources, i.e. some knowledge about their flux
distribution, in the frame of a Bayesian
formalism~\citep{hob03,psnakesI}.

With the previous single frequency methods and for the case of the
\emph{Planck} mission we expect to reach detection limit fluxes that
range from a few hundred mJy to several Jy, depending on the
frequency channel, and to obtain catalogues that will contain from
several hundreds to a few thousand
objects~\citep{can06,challenge08short}. This is satisfactory, but we feel
that we could do better if we were able to use multi-wavelength
information in some way. 

For now, multi-wavelength detection of EPS in CMB images remains a
largely unexplored field. In recent years, some attempts have been
done in this direction~\citep{naselsky02b,chen08,wright09short}. More
recently,~\citet{herranz08a} have introduced the technique of
\emph{`matched matrix filters'} (MTXF) as the first fully
multi-frequency, non-parametric, linear filtering technique that is
able to find EPS and to do unbiased estimations of their fluxes thanks
to the distinctive spatial behaviour of the sources, while at the same
time does incorporate some multi-wavelength information, without
assuming any specific spectral behaviour for the
sources.~\citet{mtxf09} have applied the MTXF to realistic simulations
of the \emph{Planck} radio channels, showing that it is possible to
practically double the number of detections, for a fixed reliability
level, for some of the channels with respect to the single frequency
matched filter approach.

MTXF use multi-wavelength information in such a way that it is not
necessary to make any assumption about the spectral behaviour of the
sources. In fact, in that formalism their spectral behaviour is
entirely irrelevant. All the multi-wavelength considerations concern
only to the noise\footnote{Here the term `noise' refers to all the
components except for the EPS themselves, including CMB and Galactic
foregrounds.} and its correlations. In this sense, MTXF deal only with
half of the problem. This has its advantages in terms of robustness
and reliability, but one could wish to have a technique that uses
multi-wavelength information in the modelling of \emph{both} the
signal (EPS) and the noise. But, as we mentioned before, the spectral
behaviour of the EPS is not known.

In this paper we will show that even if the spectral behaviour of the
EPS is unknown a priori, it is still possible to determine it directly
from the data by means of an adaptive filtering scheme that
incorporates multi-frequency information not only through the noise
correlations among channels, but also about the sources themselves.

The problem is, in more than one sense, similar to the problem of
detecting SZ clusters (that is the reason we discussed that case a few
paragraphs above). In the SZ case the spectral behaviour of the
sources is known, but not their size. A way to deal with this is to
introduce the scale of the source as a free parameter (for example,
the cluster core radius $r_c$) in the design of a \emph{`matched
multifilter'} (MMF) and to optimise the value of this parameter for
each source so that a maximum signal to noise ratio is obtained after
filtering~\citep[see the details
in][]{herr02a,herr02b,herr05,schaf06,melin06}. As the problem depends
on the optimisation of one single parameter, the method is easy to
implement in codes that are relatively fast.

In this paper we introduce a modification of the MMF technique in
which the mixing coefficients of the frequency dependence vector of
the sources are considered as free parameters to be optimised. As we
will show, if the number of frequency channels is $N$ and we choose
wisely a fiducial frequency of reference, the number of free
parameters to optimise is $N-1$. For simplicity, throughout this paper
we will use as an example the case of two channels, $N=2$, but the
method is valid for any number of channels $N>1$. 

The structure of this paper is as follows. In section~\ref{sec:metodo}
we will summarise the formulae of the MMF and we will introduce the
modification that allows us to use it for detecting EPS in arbitrarily
chosen pairs of frequency channels. We will show that the filter can
be in this case factorised in such a way that a particularly fast
implementation of the method can be achieved.  In
section~\ref{sec:simulaciones} we will describe the simulations that
we use to test the method.  The results of the exercise will be
described in section~\ref{sec:resultados}. Finally, in
section~\ref{sec:conclusiones} we will draw some conclusions.

\section[]{METHOD} \label{sec:metodo}

\subsection{The single frequency approach} \label{sec:single}

Let us assume a set of images corresponding to the same area of the
sky observed simultaneously at $N$ different frequencies:
\begin{equation} \label{modelo}
  y_{\nu}({\bf x})=f_{\nu} s_{\nu}(\mathbf{x})+n_{\nu}({\bf x}),
\end{equation}
\noindent 
where $\nu=1,\ldots,N$. At each frequency $\nu$, $y_{\nu}$ is the
total signal in the pixel ${\bf x}$ and $s_{\nu}$ represents the
contribution of the point source to the total signal $y_{\nu}$; for
simplicity let us assume there is only one point source centered at
the origin of the image; $f_{\nu}$ is the frequency dependence of the
point source; and $n_{\nu}$ is the \emph{background} or generalized
noise (containing not only the instrumental noise, but also the
contributions of the rest of components).

The intrinsic angular size of the point sources is smaller than the
angular resolution of the detector. At each observing frequency, each
source is convolved with the corresponding antenna beam. For
simplicity we will assume the antenna beam can be well described by a
symmetric 2D Gaussian function. Then we can write
\begin{equation} \label{eq:amplitud}
s_{\nu}(x)=A \tau_{\nu}(x),
\end{equation}
\noindent
where $x=|{\bf x}|$ (since we are considering symmetric beams), $A$ is
the \emph{amplitude} of the source and $\tau$ is the spatial template
or \emph{profile}. The background $n_{\nu}({\bf x})$ is modelled as a
homogeneous and isotropic random field with average value equal to
zero and power spectrum $P_{\nu}$ defined by
\begin{equation} \label{espectro_potencias}
  \langle n_{\nu}({\bf q})n^*_{\nu}({\bf q'}) \rangle =
  P_{\nu}\delta_D^2({\bf q}-{\bf q'}),
\end{equation}
\noindent
where $n_{\nu}({\bf q})$ is the Fourier transform of $n_{\nu}({\bf x})$
and $\delta_D^2$ is the 2D Dirac distribution.

In the single frequency approach each channel is processed separately
and independently from the other frequency channels. This approach is
robust in the sense that it is not necessary to assume anything about
the spectral behaviour of the sources. The main drawback of the single
frequency approach, however, is that one misses the potential noise
reduction that could be obtained with a wise use of the information
present at the other frequencies.

The standard single frequency point source detection method in the
literature is based on the \emph{matched
  filter}~\citep{tegmark98,barr03,can06}. The matched filter is the
optimal linear detector for a single map in the sense that it gives
the maximum signal to noise amplification. The matched filter can be
expressed in Fourier space in the following way:
\begin{equation} \label{eq:MF}
\psi_{MF}(q)  =  \frac{\tau(q)}{a P(q)}, \ \
a            =  \int d \mathbf{q} \frac{\tau^2(q)}{P(q)}.  
\end{equation}
\noindent
Here $a$ is a normalisation factor that preserves the source amplitude
after filtering.

\subsection[]{The Matched Multifilter} \label{sec:mmf}

In the multi-frequency approach we take into account the statistical
correlation of the noise between different frequency channels and the
frequency dependence of the sources. Now let us model the background
$n_{\nu}({\bf x})$ as a homogeneous and isotropic random field with
average value equal to zero and crosspower spectrum $P_{\nu_1\nu_2}$
defined by:
\begin{equation} \label{cross_espectro_potencias}
  \langle n_{\nu_1}({\bf q})n^*_{\nu_2}({\bf q'}) \rangle =
  P_{\nu_1\nu_2}\delta_D^2({\bf q}-{\bf q'})
\end{equation}
\noindent
where $n_{\nu}({\bf q})$ is the Fourier transform of $n_{\nu}({\bf
  x})$ and $\delta_D^2$ is the 2D Dirac distribution. Let us define a
set of $N$ linear filters $\psi_{\nu}$ that are applied to the data
\begin{equation}   \label{mapa_filtrado}
  w_{\nu}({\bf b})=\int{d{\bf x} \ y_{\nu}({\bf x})\psi_{\nu}({\bf
      x};{\bf b})} = \int{d{\bf q} \ e^{-i{\bf q}\cdot{\bf
      b}}y_{\nu}({\bf q})\psi_{\nu}(q)}.
\end{equation}
\noindent
Here ${\bf b}$ defines a translation.  The right part of equation
(\ref{mapa_filtrado}) shows the filtering in Fourier space, where
$y_{\nu}({\bf q})$ and $\psi_{\nu}(q)$ are the Fourier transforms of
$y_{\nu}({\bf x})$ and $\psi_{\nu}({\bf x})$, respectively. The
quantity $w_{\nu}({\bf b})$ is the the filtered map $\nu$ at the
position $\mathbf{b}$. The \emph{total filtered map} is the sum
\begin{equation} \label{eq:total_filtered_map}
  w({\bf b})=\sum_{\nu}{w_{\nu}({\bf b})}.
\end{equation}
Therefore, the total filtered field is the result of two steps: a)
filtering and b) fusion. During the first step each map $y_{\nu}$ is
filtered with a linear filter $\psi_{\nu}$; during the second step the
resulting filtered maps $w_{\nu}$ are combined so that the signal $s$
is boosted while the noise tends to cancel out.  Note that the fusion
in eq. (\ref{eq:total_filtered_map}) is completely general, since any
summation coefficients different than one can be absorbed in the
definition of the filters $\psi_{\nu}$. Then the problem consists in
how to find the filters $\psi_{\nu}$ so that the total filtered field
is \emph{optimal} for the detection of point sources.

The total filtered field $w$ is \emph{optimal} for the detection of
the sources if
\begin{enumerate}
\item $w({\bf 0})$ is an \emph{unbiased} estimator of the amplitude of
  the source, so $\langle w({\bf 0})\rangle=A$;
\item the variance of $w({\bf b})$ is minimum, that is, it is an
  \emph{efficient} estimator of the amplitude of the source.
\end{enumerate}
\noindent
If the profiles $\tau_{\nu}$ and the frequency dependence $f_{\nu}$
are known and if the crosspower spectrum is known or can be estimated
from the data, the solution to the problem is already known: the
matched multifilter~\citep[MMF,][]{herr02a}:
\begin{equation} \label{eq:mmf}
 \mathbf{\Psi}(q)=\alpha \ {\bf P}^{-1} \mathbf{F}, \ \ 
 \alpha^{-1}=\int{d{\bf q} \  \mathbf{F}^t{\bf
     P}^{-1} \mathbf{F}},
\end{equation}
\noindent
where $\mathbf{\Psi}(q)$ is the column vector
$\mathbf{\Psi}(q)=[\psi_{\nu}(q)]$, $\mathbf{F}$ is the column vector
$\mathbf{F}=[f_{\nu}\tau_{\nu}]$ and ${\bf P}^{-1}$ is the inverse
matrix of the cross-power spectrum {\bf P}. Finally, we can obtain the
variance of the total filtered field, given by the following
expression:
\begin{equation} \label{eq:variance}
  \sigma^2_w=\int{d{\bf q}\mathbf{\Psi}^t{\bf P}\mathbf{\Psi}}=\alpha
\end{equation}

\subsection{MMF with unknown source frequency
  dependence} \label{sec:MMF_g}

As it was previously discussed, the problem is that the frequency
dependence $f_{\nu}$ of the sources is not known a priori. Then, the
possibilities are either a) to admit defeat, returning to the single
frequency approach, b) to devise a filtering method that does not use
the frequency dependence of the sources altogether or c) to model
somehow the unknown frequency dependence in the framework of some
optimisation scheme. The second approach was explored
in~\cite{herranz08a,mtxf09}. In this work we will study the third
approach to the problem.

Before addressing this problem, it will be useful to rewrite
eq.(\ref{eq:mmf}) in a slightly different way. Let us write the vector
$\mathbf{F}=[f_{\nu}\tau_{\nu}]$ in matrix form as
\begin{equation} \label{eq:factor}
\mathbf{F}=\mathbf{T}(q) \mathbf{f}(\nu),
\end{equation}
\noindent
with $\mathbf{T}$ a diagonal matrix $\mathbf{T}(q) =
\mathrm{diag}[\tau_1(q),\ldots,\tau_N(q)]$ and $\mathbf{f}=[f_{\nu}]$
the vector of frequency dependences. Note that all the dependence in
$q$ is included in the matrix $\mathbf{T}$; this fact will be useful
later.

Now imagine that $\mathbf{f}$ describes the true (unknown) frequency
dependence of the sources and that 
$\mathbf{g}=[g_{\nu}]$, $\nu=1,\ldots,N$ is a new vector, of equal
size as $\mathbf{f}$ but whose elements can take any possible
value. We can define the MMF for vector  $\mathbf{g}$
\begin{eqnarray} \label{eq:mmf_g}
 \mathbf{\Psi}_{\mathbf{g}}(q) &  = & \alpha_{\mathbf{g}} \ {\bf P}^{-1}
   \mathbf{T} \mathbf{g}, \nonumber \\
 \alpha_{\mathbf{g}}^{-1} & = & \int d\mathbf{q} \  \mathbf{g}^t 
 \mathbf{T} 
     \mathbf{P}^{-1} \mathbf{T} \mathbf{g} = 
     \mathbf{g}^t 
     \mathbf{H} 
     \mathbf{g},
\end{eqnarray}
\noindent
where $\mathbf{H}=\int d\mathbf{q} \mathbf{T} \mathbf{P}^{-1}
\mathbf{T}$ and we have used the facts that $\mathbf{T}^t =
\mathbf{T}$ and that vector $\mathbf{g}$ does not depend on
$\mathbf{q}$ and can therefore go out of the integral. When applied to
a set of images where there is present a point source with true
amplitude $A$ and true frequency dependence $\mathbf{f}$, the filters
$\mathbf{\Psi}_{\mathbf{g}}$ will lead to an estimation of the
amplitude
\begin{equation} \label{eq:Ag}
A_{\mathbf{g}} = w_{\mathbf{g}}(\mathbf{0}) =  
\alpha_{\mathbf{g}} \ A \ 
\mathbf{g}^t \mathbf{H} \mathbf{f}.
\end{equation}
\noindent
Note that if $\mathbf{g} \neq \mathbf{f}$, then $A_{\mathbf{g}} \neq
A$. On the other hand, the variance of the filtered field would be, in
analogy with eq. (\ref{eq:variance}), $\sigma_{\mathbf{g}}^2 =
\alpha_{\mathbf{g}}$. Let us finally define the signal to noise ratio of the
source in the total filtered map as
\begin{equation} \label{eq:snr}
SNR_{\mathbf{g}} = \frac{A_{\mathbf{g}}}{\sigma_{\mathbf{g}}}.
\end{equation}
\noindent
Then we can ask what is the vector $\mathbf{g}$ that maximizes the
signal to noise ratio $SNR_{\mathbf{g}}$. Intuition alone indicates
that $SNR_{\mathbf{g}}$ is maximum if and only if $\mathbf{g} =
\mathbf{f}$. This can be formally proved with little effort by taking
variations of $\mathbf{g}$.

Then the problem can be solved via a maximisation algorithm. In the
case of a non blind search, where the position of a given point source
is known, one can focus on that point source and iteratively try
values of the elements of $\mathbf{g}$ until a maximum signal to noise
is reached. In the case of a blind search, the situation is a little
bit more difficult because in a given image there may be many
different objects $s_i$ with a different solution $\mathbf{g}_i$.  A
way to proceed is to filter many times the image, using each time a
different set of values of the elements of $\mathbf{g}$ so that the
appropriate range of frequency dependences is sufficiently well
sampled, and then to proceed counting one by one all the possible
detections and associating to each one the values of $\mathbf{g}$ that
maximize the signal to noise ratio of that source in particular.

We would like to remark that this situation is very similar to the
case of the detection of galaxy clusters with unknown angular size
described in~\cite{herr02a,herr02b}. In that case the frequency
dependence $\mathbf{f}$ was known but the size of the clusters (their
source profile) was not. The cluster profile can be parametrised as a
modified beta profile with a free scale parameter $r_c$ (typically,
the cluster core radius). In~\cite{herr02a,herr02b} it was shown that
the true scale of the clusters can be determined by maximizing the
signal to noise ratio of the detected clusters as a function scale
$r_c$ of the filter.

In our case, factorisation (\ref{eq:factor}) leads to equations
(\ref{eq:mmf_g}) and (\ref{eq:Ag}); this is very convenient for
implementation of the MMF when many filtering steps are necessary. The
most time-consuming part of the filter is the calculation of matrices
$\mathbf{P}$ and $\mathbf{T}$ because they must be calculated for all
values of $\mathbf{q}$. In the case of clusters with unknown size
$\mathbf{T}$ had to be calculated for every value of $r_c$. However,
in the case we are considering in this paper the only quantity that
varies during the maximisation process are the elements of vector
$\mathbf{g}$. This allows us to compute the integrals of matrix
$\mathbf{H}$ only once for each set of images. As a result, applying
the MMF to large numbers of point sources with unknown frequency
dependence is, in general, much faster than applying the MMF to the
same number of clusters with unknown source profile.

The main difference is that while in the case of galaxy clusters it
was necessary to maximize with respect to only one single parameter
(the core radius), in the case of the unknown frequency dependence it
is necessary to maximize with respect to the $N$ components of vector
$\mathbf{g}$. This procedure can require a very large number of
computations if $N$ is big. Although we have just seen that each free
parameter of the frequency dependence can be mapped much faster than
each free parameter of the source profile, we are still interested in
reducing the number of computations as much as possible.

\subsection{Number of degrees of freedom of vector $\mathbf{g}$}

For $N$ images, vector $\mathbf{g}=[g_1,\ldots,g_N]$ has $N$ degrees
of freedom. This makes the optimisation procedure more complex and
computationally expensive. The situation can be lightened if we choose
one of the frequencies under consideration as our fiducial frequency
of reference. Let us choose for example a concrete frequency $j \in
\{1,\ldots,N\}$ to be our fiducial frequency of reference, then
\begin{equation}
\langle y_j(0) \rangle = A  f_j \tau_j(0) = A
\end{equation}
\noindent 
and therefore, since the profile $\tau_j$ is normalised to unity,
$f_j$ must be equal to one. Therefore, we must look for vectors
$\mathbf{g}= {g_1,\ldots,g_{j-1},1,g_{j+1},\ldots,N}$ and the number
of independent degrees of freedom is $N-1$. If the number of channels
is $N=2$, there is only one degree of freedom for the optimisation
problem. In the next sections of this paper we will consider, for
simplicity, the case of two channels, but we would like to remark that
the extension to $N>2$ frequencies is straightforward.

\subsection{Optional parametrisation of vector $\mathbf{g}$}

Another way to reduce the number of degrees of freedom of vector
$\mathbf{g}$ is to find a suitable parametrisation for it. For
example, the power law relationship
\begin{equation} \label{eq:flux_powlaw}
I(\nu) = I_0 \left( \frac{\nu}{\nu_0} \right)^{-\gamma},
\end{equation}
where $I(\nu)$ is the flux at frequency $\nu$, $\nu_0$ is a frequency
of reference, $I_0$ is the flux at that frequency of reference and
$\gamma$ is the \emph{spectral index}, is widely used in the
literature. If eq. (\ref{eq:flux_powlaw}), then the reference flux
$I_0$ can easily be related to the reference amplitude $A$ of the
sources and the number of degrees of freedom is just one, the spectral
index $\gamma$.

However, the parametrisation of vector $\mathbf{g}$ has its own
risks. For example, it is known that eq. (\ref{eq:flux_powlaw}) is
valid only as an approximation for any given frequency interval and
that its validity decreases as the size of the interval grows. If we
choose to follow the parametric approach we know for sure that the
results will be less and less accurate as the number $N$ of channels
grows, especially if the separation between frequency channels is
large. On the other hand, eq. (\ref{eq:flux_powlaw}) is alway exact if
$N=2$. Therefore we can safely use the parametrisation
(\ref{eq:flux_powlaw}) for $N=2$ channels, without loss of
generality. Since the number of degrees of freedom is one either if
(\ref{eq:flux_powlaw}) is used or not, the use of the parametrisation
is irrelevant in this case. However, we may be interested in using it
for historical, didactic and practical motivations. For example,
eq. (\ref{eq:flux_powlaw}) is useful to express the physical
properties of the sources in terms of their (steep, flat, inverted,
etc.) spectral index.

We would like to remark that frequency dependence parametrisation, in
the form of eq.  (\ref{eq:flux_powlaw}) or any else other way, may or
not be useful in some cases, but \emph{it is not essential} at all for
the method we propose in this paper.

\section{Simulations} \label{sec:simulaciones}

In order to illustrate the MMF method described above and to compare
the MMF multi-frequency approach with the single frequency approach,
we have performed a set of basic, yet realistic, simulations. We
consider the case of two frequency channels ($N=2$). Generalisation to
more frequency channels is possible, as discussed in
section~\ref{sec:MMF_g}, but we choose to keep things simple in this
paper.

For this example we take the case of the \emph{Planck}
mission~\citep{planck_tauber05}. We will consider the 44 GHz and 100
GHz \emph{Planck} channels. The choice of the pair channels is not
essential: any other pair of channels would have served the same for
this exercise. This particular choice allows to study the case of
radio sources in two not adjoining channels with different
instrumental settings: the 44 GHz channel belongs to the Low Frequency
Instrument of \emph{Planck} and the 100 GHz channel belongs to the
High Frequency Instrument.

For the simulations we have used the \emph{Planck} Sky
Model\footnote{http://www.apc.univ-paris7.fr/APC\_CS/Recherche/Adamis/
PSM/psky-en.php}~\citep[PSM,][in preparation]{psm}, a flexible
software package developed by \emph{Planck} WG2 for making
predictions, simulations and constrained realisations of the microwave
sky. The simulated data used here are the same as
in~\citet{challenge08short}, where the characteristics of the
simulations are explained in more detail. Maps are expressed in
$(\Delta T/T)$, thermodynamic units. Simulations include all the
relevant astrophysical components: the CMB sky is based on a Gaussian
realisation assuming the WMAP best-fit $C_{\ell}$ at higher
multipoles; Galactic emission is described by a three component model
of the interstellar medium comprising free-free, synchrotron and dust
emissions.  Free-free emission is based on the model of~\citet{dick03}
assuming an electronic temperature of 7000 K. The spatial structure of
the emission is estimated using a H$\alpha$ template corrected for
dust extinction. Synchrotron emission is based on an extrapolation of
the 408 MHz map of~\citet{has82} from which an estimate of the
free-free emission was removed.  A limitation of this approach is that
this synchrotron model also contains any dust anomalous emission seen
by WMAP at 23 GHz. The thermal emission from interstellar dust is
estimated using model 7 of~\citet{fink99}.

For the purely descriptive purposes of this example, we take eight
different regions of the sky located at intermediate Galactic latitude
(four of them uniformly distributed across the 40$^{\circ}$ North
Galactic latitude parallel and four of them distributed in the same
way 40$^{\circ}$ South of the Galactic plane). For each region we
select a $512\times 512$ pixel square patch (at 44 and 100 GHz). Pixel
size is 1.72 arcmin for the two frequencies. Therefore, each patch
covers an area of 14.656 square degrees of the sky. When both patches
have been selected, we add simulated extragalactic point sources with
a spectral behaviour described by eq. (\ref{eq:flux_powlaw}). We take
as frequency of reference $\nu_0 = 44$ GHz. Note that
eq. (\ref{eq:flux_powlaw}) is expressed in flux units and the maps are
in $(\Delta T/T)_{th}$: we make the appropriate unit conversion before
adding the sources. The antenna beam is also taken into account: the
full width at half maximum is FWHM=24 arcmin for the 44 GHz channel
and FWHM=9.5 arcmin at 100 GHz. Finally, after doing that, we have
added to each patch uniform white noise with the nominal levels
specified for \emph{Planck}~\citep{bluebook} and this pixel size.

We are interested in comparing the performance of the multi-frequency
approach with that of the single-frequency matched filter. In
particular, we expect to be able to detect fainter sources with the
MMF than with the MF. From recent works~\citep{can06,challenge08short}
we know that in this kind of \emph{Planck} simulations, the MF can
detect sources down to fluxes $\sim 0.3$ Jy (the particular value
depends on the channel and the region of the sky). Here we will
simulate sources in the interval $[0.1,1.0]$ Jy plus a few cases, that
will be described below, where even lower fluxes are
necessary. Regarding the spectral index of the sources, according
to~\citet{gnuevo08}, most radio galaxies observed by WMAP at fluxes
$\sim 1$ Jy show spectral indices that lie in the range $(-1.0,1.4)$.

We sample the interesting intervals of flux and spectral index by
simulating sources with fluxes at 44 GHz $I_0 = \{ 0.1, 0.2, 0.3, 0.4,
0.5, 0.6, 0.7, 0.8, 0.9, 1.0\}$ Jy and spectral indices $\gamma =
\{-1.0, -0.7, -0.4, -0.1, 0.2, 0.5, 0.8, 1.1, 1.4\}$. For each pair of
values $(I_0, \gamma)$ we have simulated 100 point sources. The point
sources are randomly distributed in the maps (obviously, the same
source is placed in the same pixel in both frequencies), with only one
constraint: it is forbidden to place a source closer than
$FWHM_{44}/2$ pixels from any other. In this way we avoid source
overlapping. Image borders are also avoided. For each set of 100
sources we proceed in the following way: we randomly choose one among
the eight patches we have and place 10 sources in it. Then we randomly
choose other patch (allowing repetition) and place the next 10
sources, and so on. In total, we have simulated 9000 sources for this
exercise (the additional simulations at fluxes below 0.1 Jy are not
included).

\section{RESULTS AND DISCUSSION}\label{sec:resultados}

In order to compare the \emph{matched filter} and the \emph{matched
  multifilter}, we use the same maps with both methods. It means that
  not only the maps, but the sources are identical for the two filters
  (their intrinsic fluxes and positions). Each simulation is filtered
  separately with the matched filter (\ref{eq:MF}) and the matched
  multifilter (\ref{eq:mmf}). In order to have a better estimation of
  the power spectra, avoiding as much as possible aliasing effects, we
  implement a power spectrum estimator that uses the 2D Hann
  window~\citep{ventana_Hann}.

We will compare performance of the two methods in terms of the
following aspects: spectral index estimation, source detection and
flux estimation.

\subsection{Source detection}

Direct comparison of source detection between the MF and the MMF is
not an obvious task because for $N=2$ input images the MF produces two
filtered images, whereas the MMF produces only one combined filtered
map. Whereas the meaning of `detection' in the latter case is
straightforward (once a detection criterion is chosen, a source is
detected or not), in the former the situation is not so clear. Imagine
we decide to apply the same detection criterion to the two MF filtered
images (which is not an obvious option), then for a given source we
can have three different outcomes of the detection:
\begin{itemize}
  \item The source may be detected in both maps.
  \item The source may be detected in only one of the maps.
  \item The source may be detected in none of the maps.
\end{itemize}
In the first two cases we can obtain point source catalogues (with
different number of objects, in principle, for the two frequencies),
but only in the first case are we able to estimate the flux at the two
frequencies and therefore the spectral index. In the case of the MMF,
if the source is detected we automatically know the spectral index and
we can use eq. (\ref{eq:flux_powlaw}) to give the fluxes at the two
frequencies.

Therefore, in the following we will distinguish two different cases
when we speak about detections with the MF. On the one hand, the
\emph{intersection} of the detections in the two channels gives us the
objects that can be used for studying the spectral index distribution;
on the other hand, the \emph{union} of the two sets gives us the total
number of objects that can be detected in, at least, one of the
channels. For the MMF, both sets are the same by definition.

Regarding the detection criterion, for simplicity we will apply the
same criterion to all the filtered maps: the widespread $5\sigma$
threshold. Note that the 5$\sigma$ threshold corresponds to different
flux values for different filters. However, in this paper we will
follow the standard 5$\sigma$ criterion for simplicity.

\begin{figure*}
\begin{center}
\includegraphics[height=19.0cm,width=22.0cm]{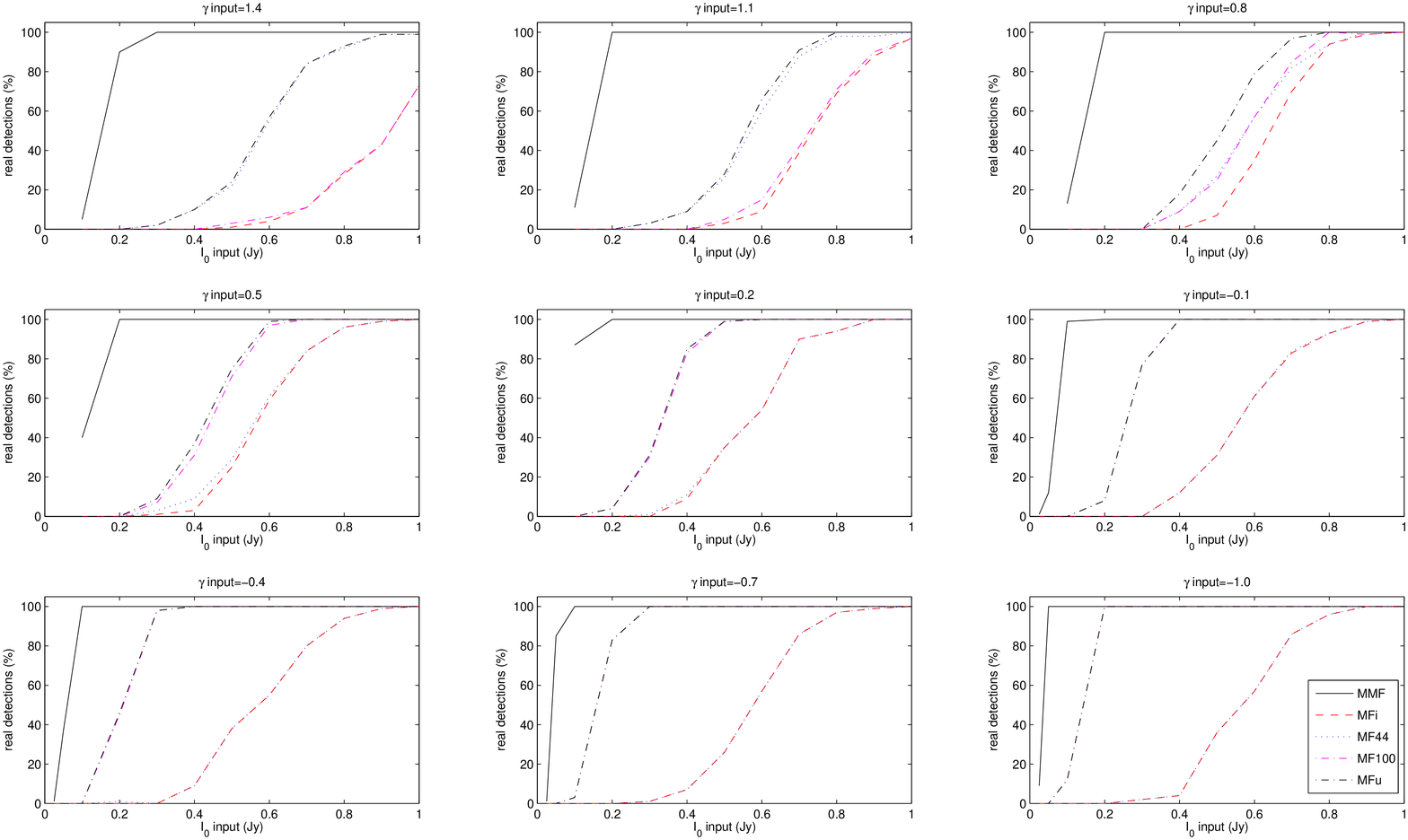}
\caption{Number of detections against the input value of $I_0$ for
  different values of the spectral index $\gamma$. MF44 represents the
  sources detected with the matched filter at 44 GHz. MF100 the same
  but at 100 GHz. MFi is the intersection of MF44 and MF100, and MFu
  is the union of MF44 and MF100. \label{detecciones}}
\end{center}

\end{figure*}

Figure~\ref{detecciones} shows the real sources (in \%) that we detect
above a 5$\sigma$ level detection whose intrinsic fluxes (values
introduced by us in the simulations) in the reference frequency
($I_0$) are the corresponding values in the horizontal
axis. Table~\ref{flujos_umbral} shows the flux at which we are able to
detect, at least, the 95\% of the sources. We can observe several
interesting aspects. The first one is the fact that the matched
multifilter improves the level of detection with respect to the
matched filter level for all the values of $\gamma$ we have inserted.

The second one is a natural selection effect: we detect more
flat/inverted sources ($\gamma \sim 0$/ negative values of $\gamma$)
at low fluxes than steep ones (positive values of
$\gamma$). Figure~\ref{detecciones} and Table~\ref{flujos_umbral} show
us that the level of detections is higher for negative values of
$\gamma$ than for positive values. Keeping in mind that the reference
frequency $\nu_0$ is equal to 44 GHz, and according to
eq. (\ref{eq:flux_powlaw}), it can be seen that for $\gamma>0$, the
simulated sources satisfy this condition: $I_{100}<I_{44}$. In this
case, the sources appear less bright at 100 GHz. In these conditions
it is quite difficult to detect sources at 100 GHz. Therefore, it
means that we are not able to give the spectral indices of these point
sources by means of the matched filter method when $\gamma$ is
strongly positive (for instance, $\gamma\gtrsim 1$). Obviously, the
smaller $\gamma$ is, the better the detection is with the matched
filter at 100 GHz. Therefore, we add from $\gamma=-0.1$ two additional
bins at $I_0=25, 50$ mJy.

Another aspect we have to remark is the similar aspect of the
detection curve for the matched filter at 44 GHz for all the $\gamma$
values (see Figure~\ref{detecciones}). The reason of this similarity
is that the reference frequency is 44 GHz, and the maps we have
simulated at this frequency are the same, independently of $\gamma$,
with only one exception: the position of the sources. This means that
statistically are equivalent (with the inherent fluctuations due to
the variation in the positions of the sources). We find a similar
number of detections by means of the matched filter at 44 GHz,
independently of $\gamma$. For this reason it is difficult to
characterize the spectral behaviour of the sources for $I_0\lesssim
0.5$ Jy, because we do not have a high percentage of detected sources
at 44 GHz below that value of $I_0$.

\begin{table*}
\begin{center}
\begin{tabular}{cccccc}
\hline
$\gamma$ & $I_{0(MMF)95\%} (Jy)$ & $I_{0(MF_{44})95\%} (Jy)$ & $I_{0(MF_{100})95\%} (Jy)$ & $I_{0(MFi)95\%} (Jy)$ & $I_{0(MFu)95\%} (Jy)$ \\
1.4 & 0.3 & 0.9 & >1.0 & >1.0 & 0.9 \\
1.1 & 0.2 & 0.8 & 1.0 & 1.0 & 0.8 \\
0.8 & 0.2 & 0.9 & 1.0 & 0.9 & 0.7 \\
0.5 & 0.2 & 0.8 & 0.6 & 0.8 & 0.6 \\
0.2 & 0.2 & 0.9 & 0.5 & 0.9 & 0.5 \\
-0.1 & 0.1 & 0.9 & 0.4 & 0.9 & 0.4 \\
-0.4 & 0.1 & 0.9 & 0.3 & 0.9 & 0.3 \\
-0.7 & 0.1 & 0.8 & 0.3 & 0.8 & 0.3 \\
-1.0 & 0.05 & 0.8 & 0.2 & 0.8 & 0.2 \\
\hline
\end{tabular}
\end{center}
\caption{Fluxes in the reference frequency ($I_0$) for which we
detect, at least, the 95\% of the sources for the different filtering
methods. MMF, MF44, MF100, MFi, MFu as the Figure~\ref{detecciones}.}
\label{flujos_umbral}
\end{table*}

Additionally, we observe that the matched multifilter is capable to
detect sources whose $I_0<0.1$ Jy for $\gamma\lesssim -0.1$. It is
interesting to compare this with the matched filter, that does not
detect sources below 0.1 Jy in the conditions of this work. This is a
result that we obtain with the method presented here. It allows us to
detect point sources whose $I_0$ is too low to be detected with the
traditional matched filter.

To summarise, we can say that the MMF improves the detection
level. Specially remarkable are the cases where the sources are near
to be \emph{flat} (central row of Figure~\ref{detecciones}). At 100
GHz, the MF recovers the 100\% of the sources for $I_0\thicksim
0.4-0.6$ Jy. Meanwhile, the MMF reaches this level for $I_0\thicksim
0.1$ Jy. This particular case is really interesting in the sense that
most of the sources have this spectral behaviour.

\subsection{Spectral index estimation} \label{sec:indice_espectral}

\begin{figure*}
\begin{center}
\includegraphics[height=16.0cm,width=19.0cm] {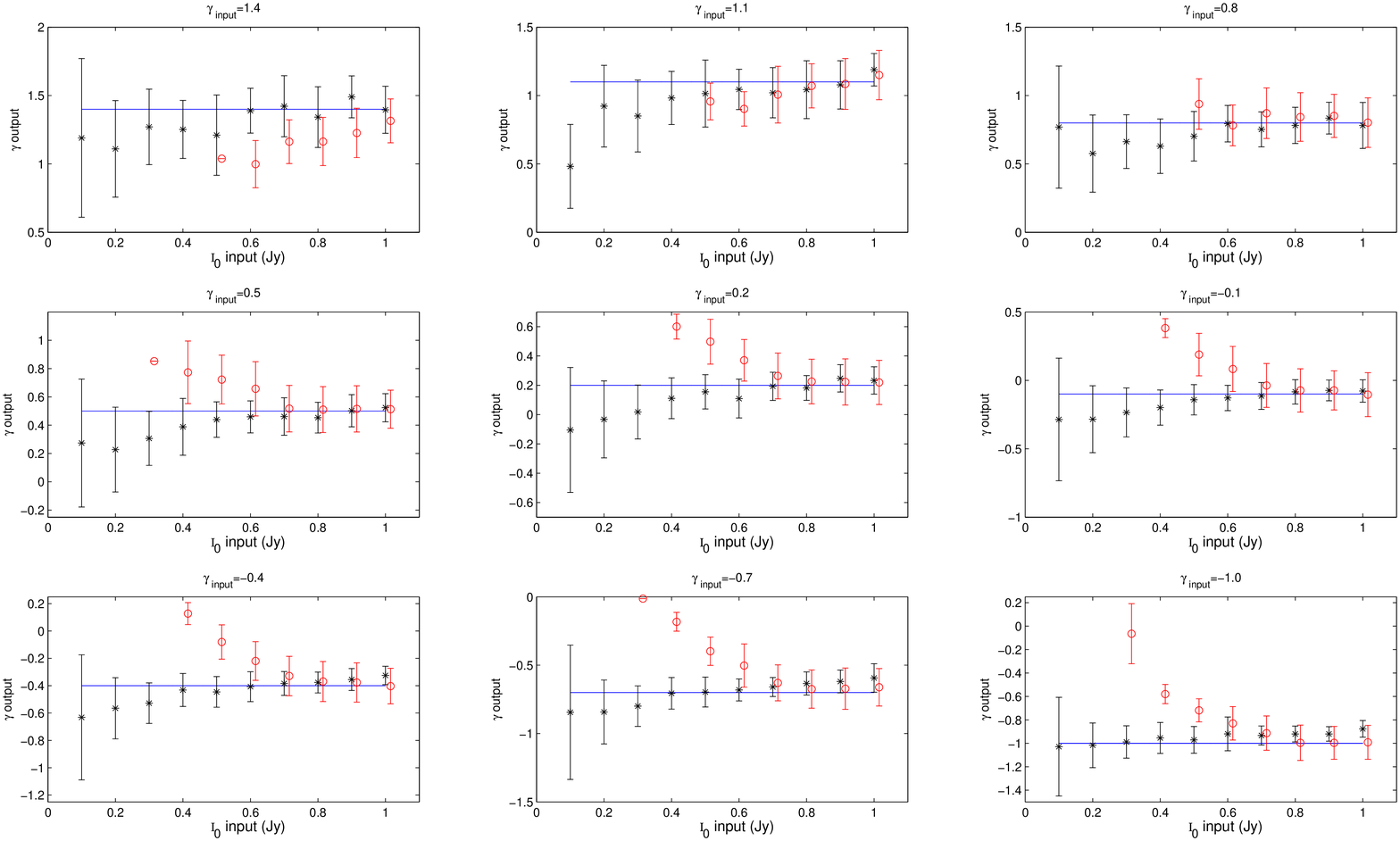}
\caption{Values of $\gamma$ recovered by means of the MMF (asterisks)
and the MF (circles). The line indicates the ideal recovering. The
circles corresponding to the MF are slightly displaced in the
horizontal axis in order to distinguish the
results. \label{gamma_estimation}}
\end{center}
\end{figure*}

As it was mentioned before, one of the quantities we want to obtain is
the spectral index of the sources (eq. \ref{eq:flux_powlaw}). As we
can see in that equation, when we know this spectral index and the
flux at the reference frequency, we are able to give an estimation of
the flux at other channels. In Figure \ref{gamma_estimation} we see
how we recover the spectral index by means of the MMF, and compare
these results with the obtained by the matched filter. In general, we
can observe that we can reach lower values of $I_0$ with the
MMF. Also, at higher values of $I_0$ than $\gtrsim 0.4$ Jy, we are
able to give, with a good degree of precision, an estimation of
$\gamma$ by means of the multifrequency method. In general, the error
bars at these values of $I_0$ are quite smaller than the bars of the
matched filter, so we recover with more accuracy the spectral index
and less uncertainty.

Other aspect is that the error bars increases when $I_0$ is
smaller. It seems logical, because we have fainter sources and a
smaller number of detections (see Figure~\ref{detecciones}). Then, at
$I_0=0.1$ Jy, we can see that the estimation of $\gamma$ is not as
good as we wish, because it has a great uncertainty. The main reason
is that the signal to noise ratio is close to the threshold level we
have imposed.

With the matched filter we estimate correctly the spectral index for
$I_0\gtrsim 0.7$ Jy at $\gamma \leq 0.8$. At higher values of $\gamma$
we find the same problem that we have mentioned before: there are not
so many detections at 100 GHz below 0.7 Jy
(Figure~\ref{detecciones}). Since the detected sources are close to
the noise level, the fluxes recovered present an overestimation with
respect to the input value due to the Eddington
bias~\citep{eddingtonBias}, an effect produced close to the noise
level by the overestimation due to the fluctuations because of the
noise in the positions where the source is located. As we said before,
and seeing the Figure~\ref{detecciones}, we detect more sources at 100
GHz than at 44 GHz for values of the spectral index smaller than 0.8.

Finally, we can observe an interesting aspect of the matched
filter. When we do not have the sufficient detections in, at least,
one channel (the sources detected are below the $\sim 40\%$ of the
total number of sources), the estimation of the spectral index is not
good. In all cases we see an overestimation of $\gamma$, except for
$\gamma={1.4,1.1}$, where we have an underestimation. This is due to
the fact that at these values of $\gamma$ the sources at 100 GHz are
fainter and the number of detections at this channel is really
small. Then, because of the Eddington bias, the flux at this frequency
is overestimated, and consequently, the value of $\gamma$ is
underestimated. The Eddington bias explains as well the overestimation
of $\gamma$ in the other cases. The only difference is that now, it is
at 44 GHz where we have a smaller number of detections. If we also see
the Figure~\ref{detecciones}, we observe that for values of
$I_0\lesssim 0.6$ Jy, we are pretty close to the noise level. It means
that the noise fluctuations in the maps produce an overestimation in
the flux at 44 GHz ($I_0$) and, in this case, an overestimation in
$\gamma$ too. Summarising, for $\gamma={1.4,1.1}$, the Eddington bias
appears at 100 GHz (underestimation of the spectral index). For the
rest of values of $\gamma$, this bias appears at 44 GHz
(overestimation of the spectral index).

\subsection{Flux estimation}

\begin{figure*}
\begin{center}
\includegraphics[height=16.0cm,width=19.0cm] {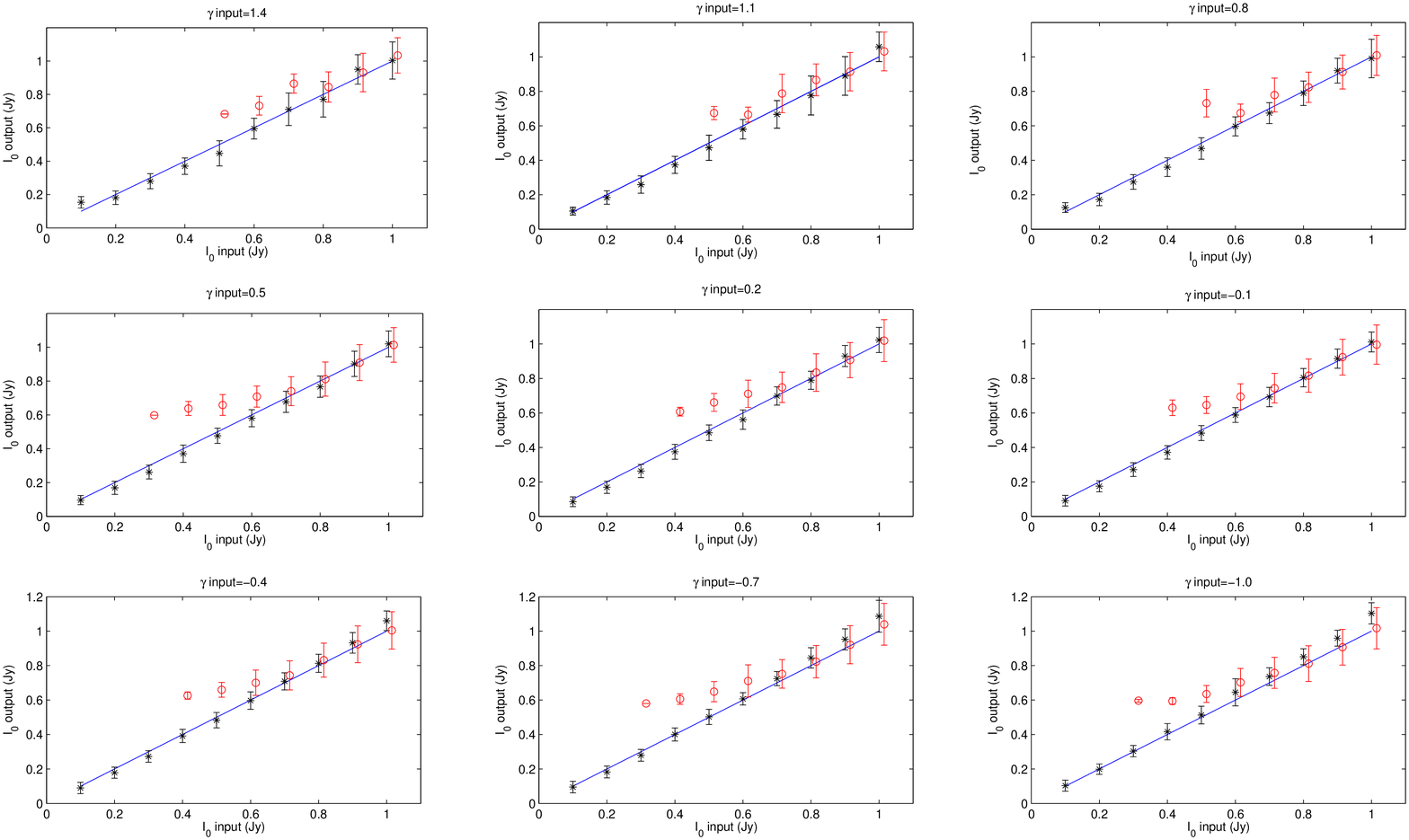}
\caption{Values of $I_0$ (flux at 44 GHz) recovered by means of the
MMF (asterisks) and the MF (circles). The line indicates the ideal
recovering. The circles corresponding to the MF are slightly displaced
in the horizontal axis in order to distinguish the
results. \label{intensidades}}
\end{center}

\end{figure*}

The other quantity that we want to recover is the flux of the source
at the reference frequency $\nu_0$. As we said in section
\ref{sec:indice_espectral}, and according to eq. \ref{eq:flux_powlaw},
when we obtain the flux at $\nu_0$, we are able to estimate the flux
at 100 GHz.

Figure~\ref{intensidades} shows the recovered flux at the reference
frequency (44 GHz) for a given value of the spectral index. The error
bars recovered with the matched filter are, in general, larger than
the ones we obtain with the matched multifilter. It is particularly
notorious at small values of $I_0$, where the recovered values of the
flux have a good agreement with respect to the input values, with
small error bars.

We observe in Figure~\ref{intensidades} the evolution of the recovered
flux. In general, for all the values of $\gamma$ that we have studied,
the matched multifilter is a suitable and effective tool to estimate
the $I_0$ of the sources. For the matched filter, we observe a good
determination of $I_0$ for input values above 0.7 Jy. For smaller
values, $I_0$ has a higher value than its real one. That is due to the
Eddington bias at 44 GHz. At this frequency, in
Figure~\ref{detecciones} we observe that for values smaller than 0.6
Jy, we only detect a $\sim 40\%$ of the total sources. That means that
many of these sources are close to this noise level. And for the
correct estimation of $I_0$ and the spectral index, we need a good
detection of the sources at the two channels. For low values of $I_0$
the number of detected objects is small and we have few statistics.

In this section we have discussed about the flux at the reference
frequency (44 GHz in our case). But we have used the matched
multifilter with two different frequencies. For this reason it is
necessary to say something related to the second channel at 100
GHz. It is important to obtain the values that we recover at 100 GHz
with the matched multifilter because the corresponding errors bars of
$I_0$ and $\gamma$ could propagate additional errors in the flux
estimation at 100 GHz (see eq. (\ref{eq:flux_powlaw})). After
extrapolating the results, we obtain the flux at 100 GHz, and compare
it with the results obtained with the matched filter. As we saw at 44
GHz, the improvement with the matched multifilter is clear respect to
the matched filter.

\subsection{Reliability}

For academic purposes, in the previous sections, we have produced the
simulations introducing 100 sources for each of the pair values of
intensity and spectral index (see section~\ref{sec:simulaciones}),
that simplifies the comparison between both filters for all the cases
under study. On the contrary, it is well known that the number of
sources per flux interval, the source number counts, is not
constant~\citep{zotti05,gnuevo08} nor the spectral index
distribution~\citep{sadler08,gnuevo08,massardi09}. In order to study
the performance of the new method under more realistic simulations we
produced a new set of simulations (100) with the following
characteristics:
\begin{itemize}
  \item We used as a background the same eight regions described in
the previous sections.
  \item The sources were simulated with an almost Poissonian
distribution (see~\citet{gnuevo05} for more details about
the method) at 44GHz, with fluxes that follow the source number counts
model of~\citet{zotti05}.
  \item The fluxes at 100GHz were estimated assuming random spectral
indices from the~\citet{gnuevo08} distribution.
  \item The point source maps were filtered with the same resolution
as the background maps and randomly added to them.
\end{itemize}
There is also another interesting quantity commonly used in the study
of the performance of a source detector: the number of spurious
sources. Spurious sources are fluctuations of the background that
satisfied the criteria of the detection method and therefore are
considered as a detected sources. It is clear that the best method
will be the one that has the best detections vs. spurious
ratio. Therefore, this time we use more realistic simulations and
count the spurious and real sources: the maps are filtered using the
MMF and the MF at both frequencies, we estimate the position and
intensity of the sources above $3\sigma$ level and, by comparing with
the input source simulations, we count the number of real and spurious
sources that we are able to detect. It is necessary to change the
detection level from $5\sigma$ to $3\sigma$ in order to observe
spurious sources and make the following analysis.

\begin{figure*}
\begin{center}
  \includegraphics[width=8.0cm] {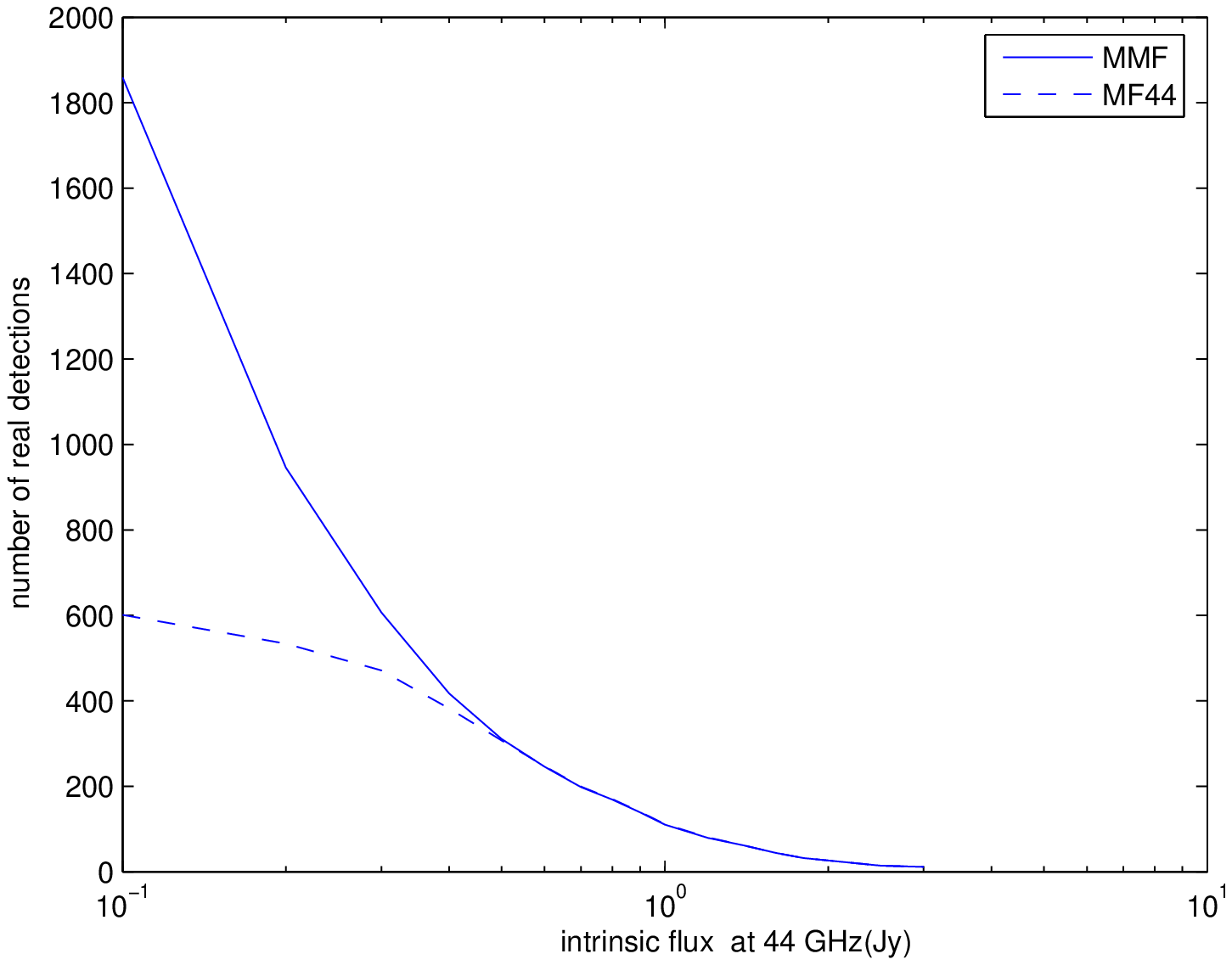}
  \includegraphics[width=8.0cm] {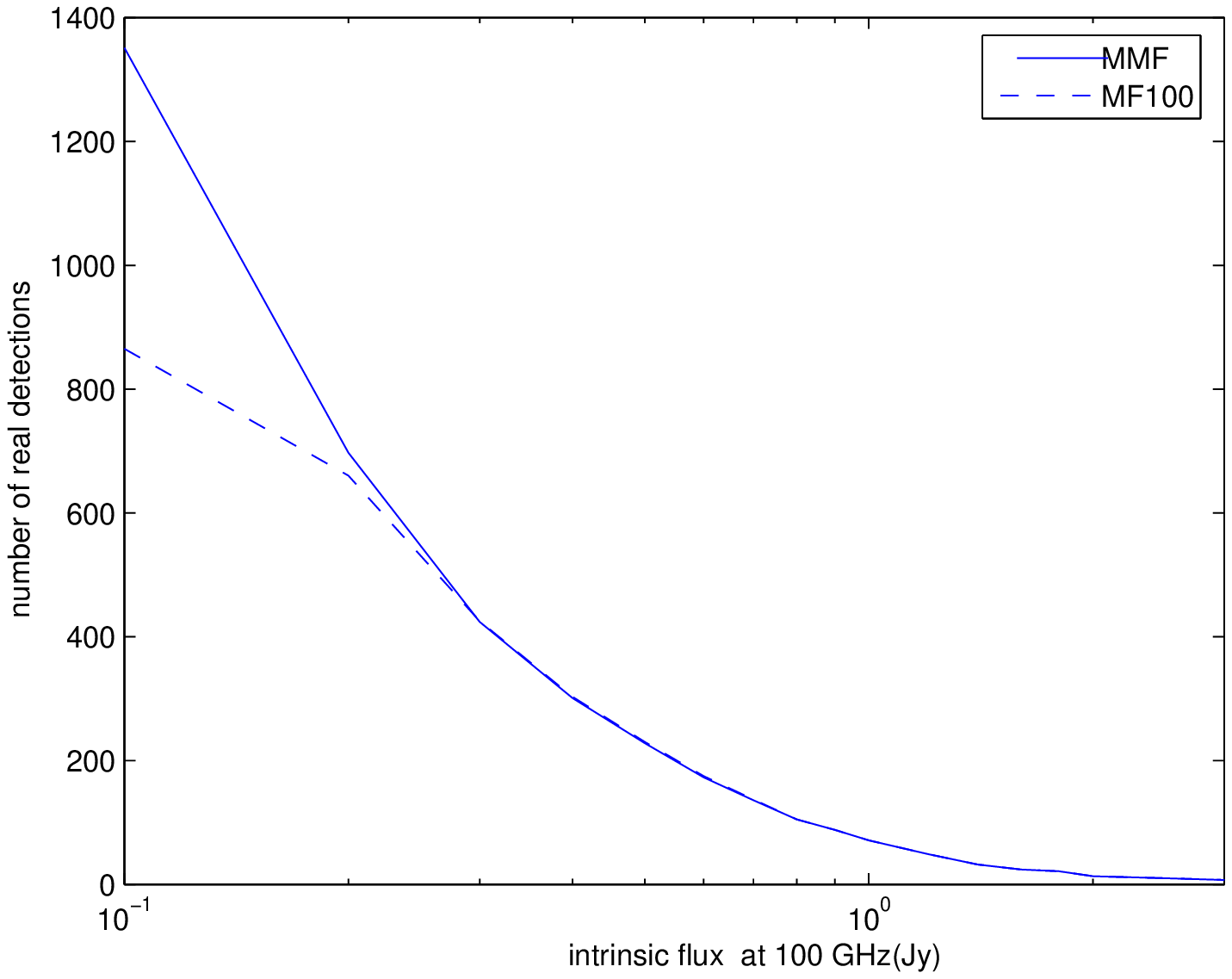}
\caption{Number of real sources recovered by the MMF (solid line) and
the MF (dashed line) at 44 GHz (left panel) and 100 GHz (right panel)
whose intrinsic fluxes are higher than the corresponding value in the
$x$ axis. \label{detec_nuevas_simul}}
\end{center}

\end{figure*}

In Figure~\ref{detec_nuevas_simul} we observe the number of real
sources that both methods are capable to detect, whose intrinsic
fluxes are higher than the corresponding value in the horizontal
axis. As we can observe at 44 GHz, MMF detects a higher number of real
sources for fluxes below $\sim 0.4-0.5$ Jy, being this difference very
important at lower fluxes. Therefore, we obtain a clear improvement
using the MMF with respect to the traditional matched filter at low
fluxes. At 100 GHz, we observe a similar behaviour, but in this case
the differences between the MF and the MMF start at $\sim 0.2$ Jy. If
we observe the Figure~\ref{detecciones}, we notice that the number of
sources detected with the MF is higher at 100 GHz than at 44 GHz for
values of the spectral index between 0 and 0.5. These values of
$\gamma$, according to the model used to simulate the point sources in
this section, are the most frequent in the real sources. This gives us
an idea about why the detection level of the MF is higher at 100 GHz.

\begin{figure*}
\begin{center}
  \includegraphics[width=8.0cm] {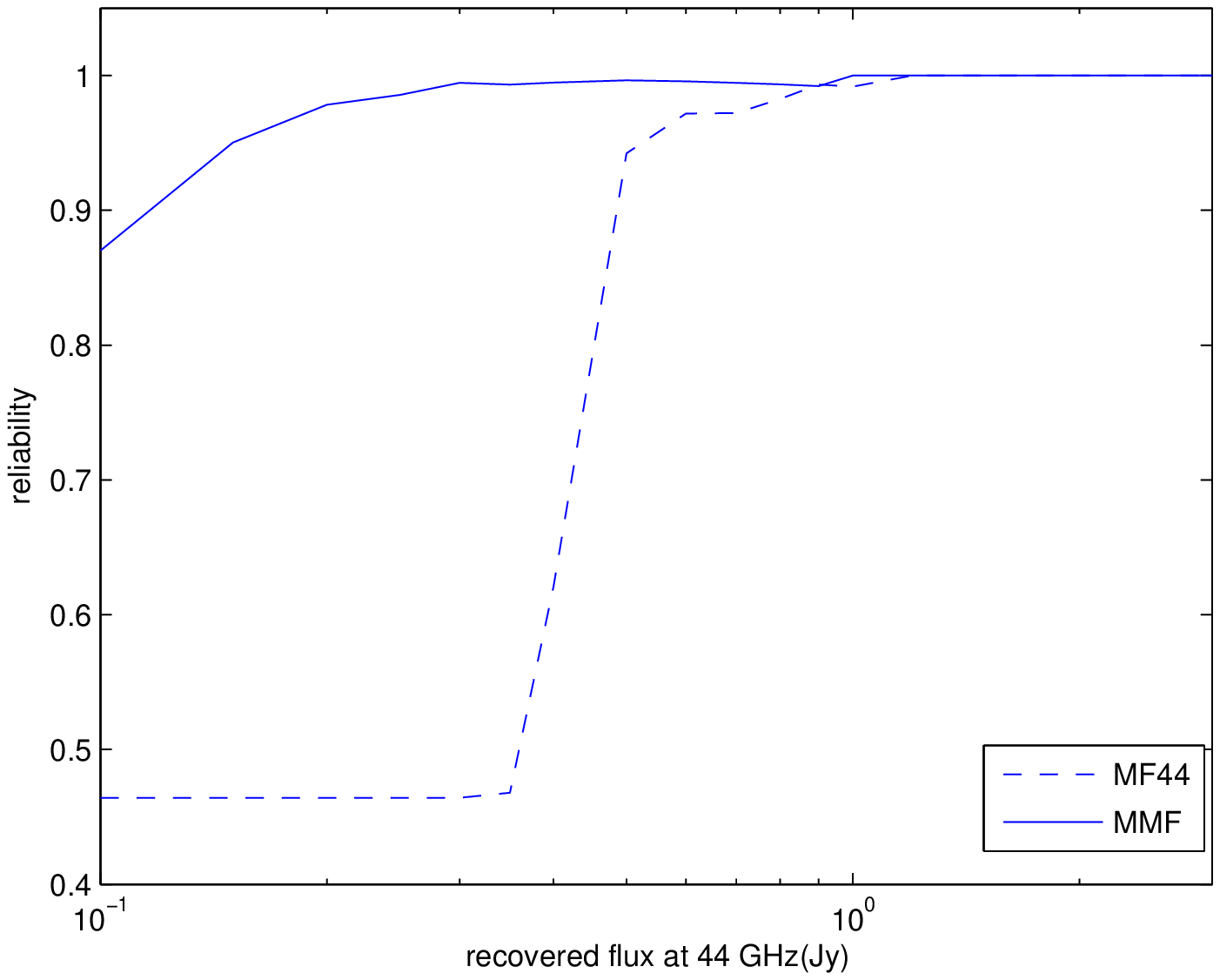}
  \includegraphics[width=8.0cm] {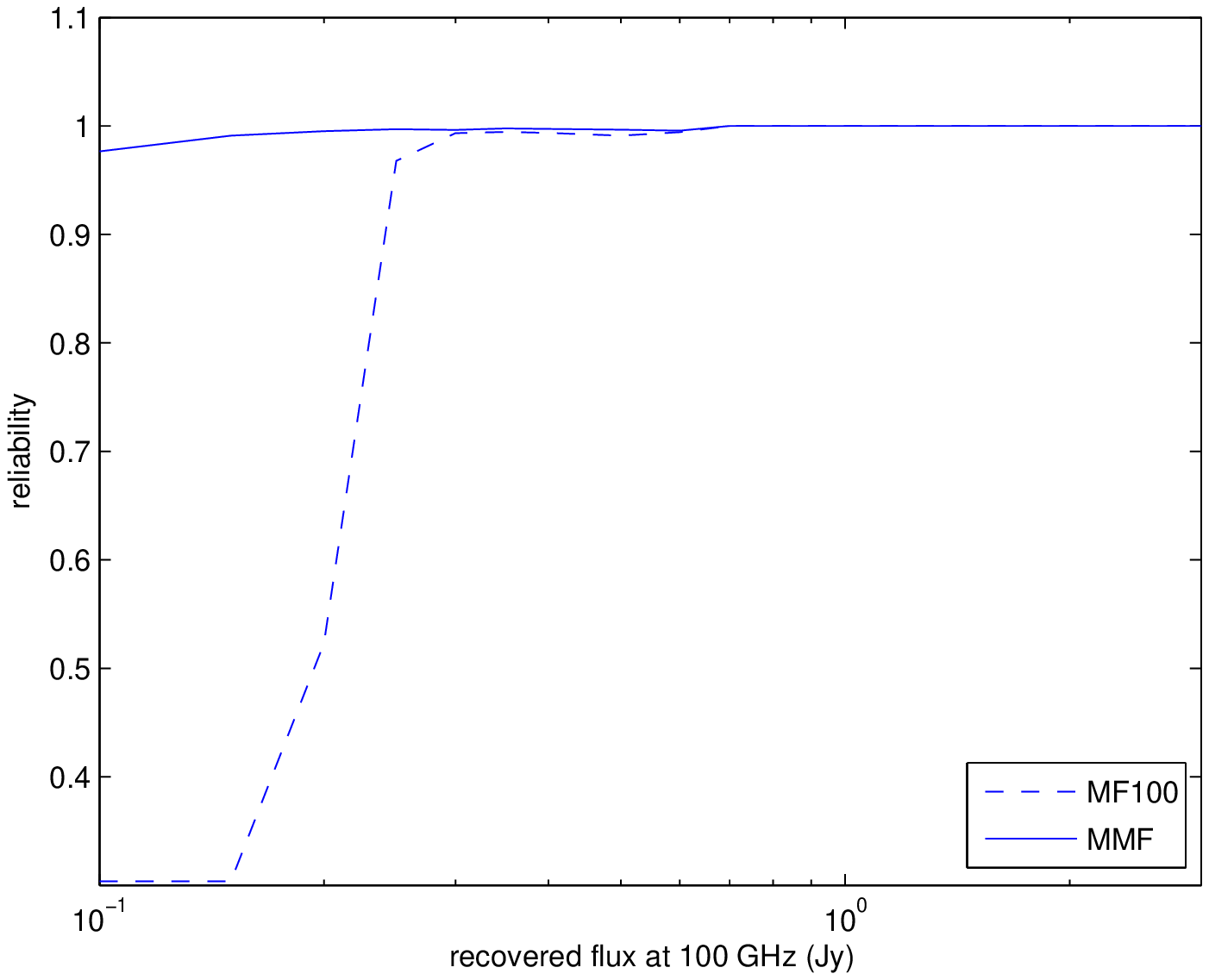}
\caption{Reliability versus recovered flux for the MMF (solid line)
and the MF (dashed line) at 44 GHz (left panel) and 100 GHz (right
panel). \label{reliab}}
\end{center}

\end{figure*}

In Figure~\ref{reliab} we compare the \emph{reliability} of both
methods at 44 and 100 GHz. Reliability above a certain recovered flux
is defined as $r=N_d/(N_d+N_s)$, where $N_d$ is the number of real
sources above that flux, and $N_s$ is the number of spurious sources
above the same flux. At 44 GHz we reach a $\sim 100\%$ of reliability
at fluxes of $\sim 0.3$ Jy. However, the MF at this frequency reaches
this level of reliability when the sources have fluxes of $\sim
0.9-1.2$ Jy. At 100 GHz we obtain better levels of reliability. For
example, with the MMF we have at 0.1 Jy more than 95\% of reliability,
and the MF reaches these values for fluxes of $\sim 0.3$ Jy. According
to the expression of the reliability, this number gives us the
percentage of real sources over the total number of sources detected
after filtering. Therefore, we can say that the MMF is more reliable
than the MF, specially at lower fluxes. Additionally, we can establish
the flux for which we have the 5\% of spurious sources. Making easy
calculations, we finally obtain that the fluxes for which we have this
percentage of spurious detections with the MF at 44 and 100 GHz are
$\sim 0.5-0.6$ Jy and $0.25$ Jy respectively. If we compare these
values with the MMF, we obtain that the fluxes are $0.15$ Jy and $<
0.1$ Jy. With these numbers one can see that the percentage of
spurious detections of the MMF is much lower than the percentage of
the MF.

\begin{figure*}
\begin{center}
  \includegraphics[width=8.0cm] {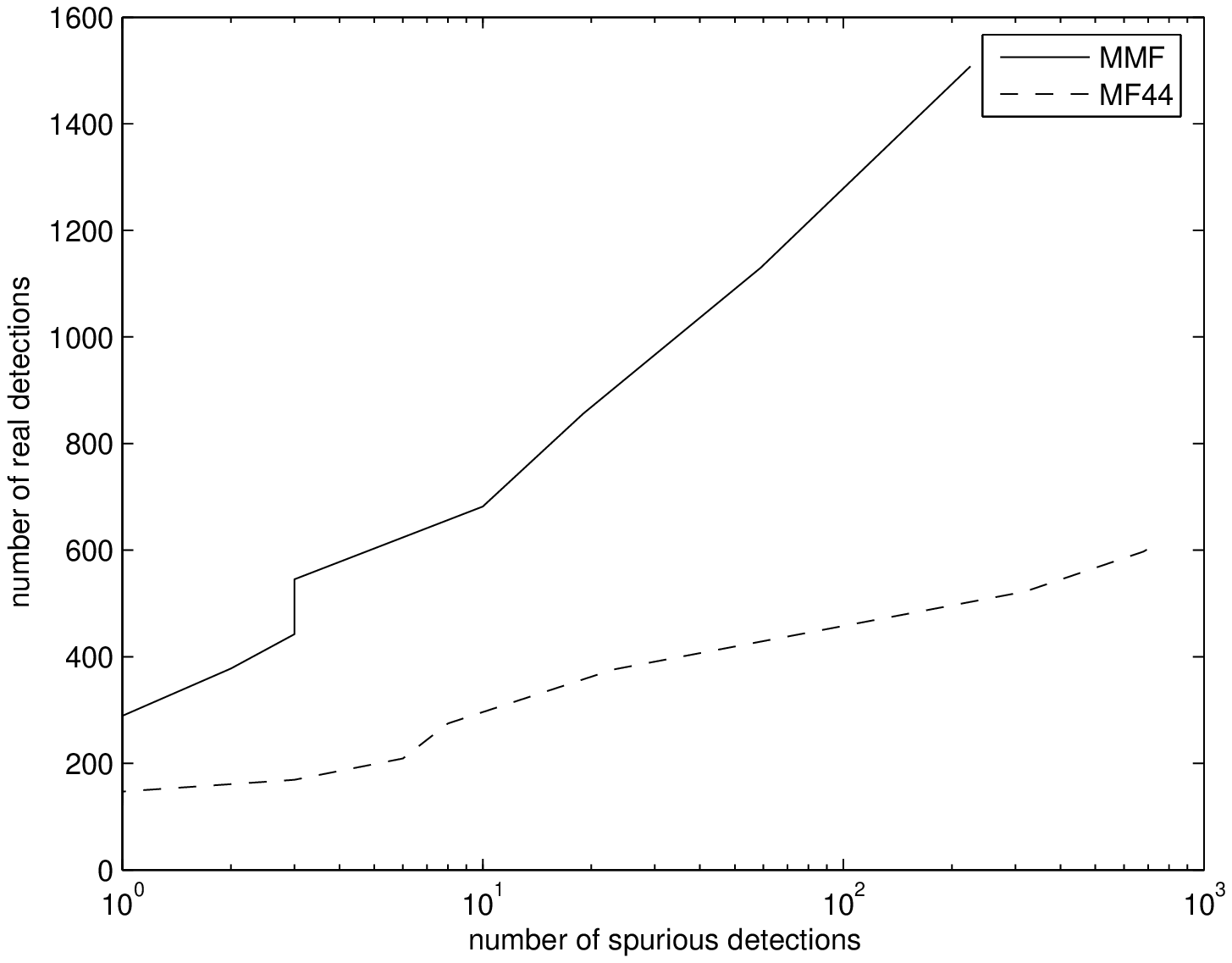}
  \includegraphics[width=8.0cm] {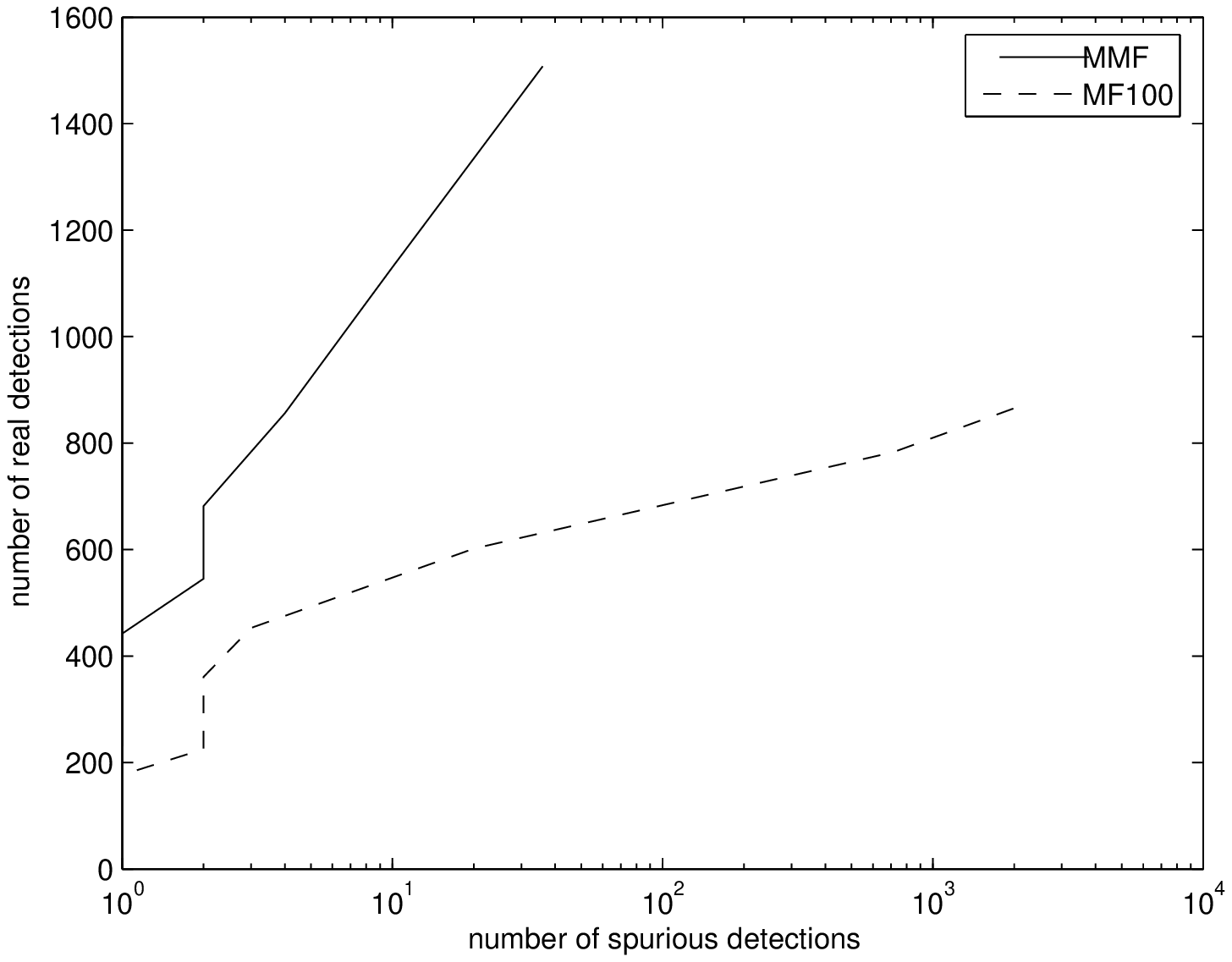}
\caption{Number of real sources recovered by the MMF (solid line) and
the MF (dashed line) at 44 GHz (left panel) and 100 GHz (right panel)
vs. the number of spurious sources. \label{ROC}}
\end{center}

\end{figure*}

Finally, we make an additional plot where we represent, for both
frequencies, the number of real sources detected vs. the number of the
spurious sources (Figure ~\ref{ROC}). In this way, what we represent
is the number of sources that a method detects given a number of
spurious sources. If we compare both plots, we can see that the curve
of the MMF is always above the MF. It means that, when we have a fixed
number of spurious detections, we detect more real sources with the
MMF.

We have to point out that the plots that we have introduced here are
not directly comparable to Figure~\ref{detecciones}. As we have seen
in this section, there are three basic and important differences:

\begin{itemize}
\item A different way to simulate the point sources.
\item A different level of detection (in this case, a $3\sigma$
level).
\item In Figure~\ref{detecciones} we represent the number of sources
with the corresponding flux in the horizontal axis. In the plots of
this section, what we represent is the number of sources whose fluxes
are higher than the corresponding value in the horizontal axis.
\end{itemize}
We can say that this new multifrequency method is better in the sense
that the number of detected sources is higher below $\sim 0.4-0.5$ Jy
and $\sim 0.2$ at 44 and 100 GHz respectively. And the reliability of
the MMF is higher for fluxes below $\sim 0.9$ Jy and $\sim 0.25$ Jy at
44 GHz and 100 GHz, respectively.

\section{Conclusions} \label{sec:conclusiones}

The detection of extragalactic point sources in CMB maps is a
challenge. One has to remove them to do a proper study of the cosmic
radiation. In addition, it is of great interest to study their
properties, spatial and spectral distributions, etc. For this reason,
we need suitable tools to detect and extract these sources. There are
many filtering techniques that have been used in this context. In this
work, we have used the matched filter, one of the most studied
techniques, and we have compared it with a new multifrequency one
based on the matched multifilter (MMF). The great difference is that
the latter takes into account information from all the channels of the
same sky region in a simultaneous way. In particular, we show an
example for $N=2$.

The different tests that we have used have shown an improvement in the
results obtained by the MMF with respect to the traditional matched
filter. The number of detections is always higher when the MMF is
used. In Figure~\ref{detecciones} we see that we have a high number of
detections with the MMF, even for small values of $I_0$. It should be
studied in more detail, but it is easy to see that one could detect
and characterize point sources with low fluxes for $\gamma<0$. For
this reason, this tool is a powerful technique to detect faint sources
in CMB maps.

Another important aspect is to give a good estimation of the
quantities that we have chosen to determine the sources, basically the
spectral index and the flux at the reference frequency. In both cases,
we can see that the MMF improves the results obtained with the matched
filter: the values are close to the input values with smaller error
bars (with one exception, the determination of the spectral index for
$I_0 \lesssim 0.5 $ Jy at positive values of the input $\gamma$). This
is a significant fact in order to be able to detect and study properly
these kind of sources.

Additionally, we have made a set of more realistic simulations in
order to study and compare both filters in the sense of the spurious
sources. We have also changed the threshold detection from 5$\sigma$
to 3$\sigma$ to find more spurious sources and make a more complete
statistical analysis. First of all, we compare the number of real
detections that we obtain with both techniques at 44 and 100
GHz. Comparing the plots of the Figure~\ref{detec_nuevas_simul}, we
appreciate that, at lower fluxes, we detect more real sources with the
matched multifilter than with matched filter. This aspect is more
notorious at 44 GHz.

One can also study the reliability of both methods. One can obtain a
high number of real detections, and simultaneously find a large number
of spurious sources. Precisely, the reliability is a quantity that
gives the number of real detections over the total number of sources
detected. Comparing the plots of the Figure~\ref{reliab}, one can
observe that the reliability of the matched multifilter is much higher
than the reliability of the matched filter for low fluxes. This
difference is particularly important at 44 GHz, where the matched
filter obtains similar values to the reliability of the matched
multifilter only for fluxes close to 1 Jy. At 100 GHz, the matched
filter reaches the reliability of the MMF at 3 Jy.

The last aspect that we use to compare both methods is to look at the
number of real sources that we have for a fixed number of spurious
detections. The most efficient method is the one that has higher
number of real detections for the same value of spurious detections. If
we see the Figure~\ref{ROC}, the best method is the MMF because its
curves are always above the MF. This means that, if we take a number
of spurious sources, the MMF recovers a larger number of real objects.

Finally, we have commented at the beginning of the
subsection~\ref{sec:MMF_g} the possibility of devising a filtering
method (the MTXF) that does not use the frequency dependence of the
sources altogether, totally independent of the frequency behaviour of
the sources (flat, steep or inverted). This fact is significant in the
sense that this filtering method is a robust technique for changes of
$f_\nu$. By contrast, it is necessary to impose the condition of
orthonormalisation of the matrix of the filters (see
\citet{herranz08a} and \citet{mtxf09} for more details). This
condition minimizes the power of the method. Meanwhile, the MMF is
more optimal in the sense of the SNR (see section~\ref{sec:MMF_g}),
but more complicated because we have to maximize another set of
parameters ($f_\nu$). As one can see, the MTXF and the MMF are
complementary.

\section*{Acknowledgements}

The authors acknowledge partial financial support from the Spanish
Ministry of Education (MEC) under project ESP2004-07067-C03-01 and the
joint CNR-CSIC research project 2006-IT-0037. LFL acknowledges the
Spanish CSIC for a JAE-Predoc fellowship and the hospitality of the
Osservatorio Astronomico di Padova (INAF) during a research
stay. Partial financial support for this research has been provided to
JLS by the Spanish MEC and to JG-N by the Italian ASI (contracts
Planck LFI Activity of Phase E2 and I/016/07/0 COFIS)and MUR. JG-N
also acknowledges a researcher position grant at the SISSA
(Trieste). ML-C acknowledges a postdoctoral fellowship from EGEE-III
(FP7 INFSO-RI 222667). The authors acknowledge the use of the Planck
Sky Model, developed by the Component Separation Working Group (WG2)
of the Planck Collaboration.

\bibliographystyle{mn2e}
\bibliography{lanz_bib}

\label{lastpage}

\end{document}